\documentclass[aps,preprintnumbers,nofootinbib,superscriptaddress,10.5pt]{revtex4}
\usepackage[dvipdfmx]{graphicx} 
\usepackage{amsmath,amssymb,amsfonts,bm,cancel} 

\def\a{\alpha} \def\b{\beta}   \def\d{\delta}  \def\e{\epsilon}   \def\th{\theta}      \def\n{\nu}               

\def\dg{\dagger} \def\del{\partial} 

\newcommand{\meV}{ {\rm meV} }   \newcommand{\MeV}{ {\rm MeV} } \newcommand{\GeV}{ {\rm GeV} } \newcommand{\TeV}{ {\rm TeV} }\newcommand{\PeV}{ {\rm PeV} }

\newcommand{\lsp}{ \left ( } \newcommand{\rsp}{ \right ) } \newcommand{\Lg}{\mathcal{L}}

\def\abs#1{\left| #1\right|}
 
  \newcommand{\Det}{{\, \rm Det \, }}

\newcommand{\Column}[3]{ \begin{pmatrix} #1 \\ #2 \\ #3 \end{pmatrix} }

\newcommand{\Diag}[3]{ \begin{pmatrix} #1 & 0 & 0 \\ 0 & #2 & 0 \\ 0 & 0 & #3 \\\end{pmatrix}}

\begin{document}


\title{\LARGE Chiral perturbative relation for neutrino masses \\  in the type-I seesaw mechanism}

\preprint{STUPP-23-262} 
\author{Masaki J. S. Yang}
\email{mjsyang@mail.saitama-u.ac.jp}
\affiliation{Department of Physics, Saitama University, 
Shimo-okubo, Sakura-ku, Saitama, 338-8570, Japan}



\begin{abstract} 

In this letter, we perform a perturbative analysis by the lightest singular value $m_{D1}$ of the Dirac mass matrix $m_{D}$ in the type-I seesaw mechanism.
A mass relation $M_{1} = m_{D1}^{2}/ |(m_{\nu})_{11}|$
is obtained for the lightest mass $M_{1}$ of the right-handed neutrino $\nu_{R1}$ and the mass matrix of the left-handed neutrinos $m_{\nu}$ in the diagonal basis of $m_{D}$. 
This relation is rather stable under renormalization because 
it is  gauge-invariant in the SM and associates with  the approximate chiral symmetry 
of $\nu_{R1}$. 
%

If diagonalization of the Yukawa matrices of leptons $Y_{\nu, e}$ has only small mixings, 
the element $(m_{\n})_{11}$ is close to the effective mass $m_{ee}$ of the neutrinoless double beta decay.  
By assuming $m_{D1} \simeq m_{u,e} \simeq 0.5$ MeV, the lightest mass is about $M_{1} \gtrsim O(100)$ TeV in the normal hierarchy and $M_{1} \sim O(10)$ TeV in the inverted hierarchy. 
Such a $\nu_{R1}$ with a tiny Yukawa coupling $y_{\nu 1} \sim O(10^{-5})$ can indirectly influence various observations.

On the other hand, the famous bound of the thermal leptogenesis $M_{1} \gtrsim 10^{9}$ GeV that requires $m_{D1} \gtrsim 30$ MeV seems to be difficult to reconcile with a simple  unified theory without a special condition.

\end{abstract} 

\maketitle

\section{Introduction}

Chiral symmetries have played an important role in the development of particle physics 
\cite{Weinberg:1968de, Witten:1978qu, tHooft:1979bh}. 
Even in the flavor physics, 
chiral symmetries are associated with small fermion masses of 
the Standard Model (SM). 
Due to the tiny singular values of the first generation, 
diagonalized Yukawa matrices $Y_{u,d,e}^{\rm diag}$ 
 have approximate chiral symmetries $U(1)_{1L} \times U(1)_{1R}$; 
\begin{align}
 R(\th_{L}) \, Y_{u,d,e}^{\rm diag} \, R (\th_{R}) \simeq Y_{u,d,e}^{\rm diag}  \, , 
\end{align}
where $R (\th) \equiv {\rm diag} (e^{i \th} \, , 1 \, , 1)$ and $\th_{L,R}$ are arbitrary real parameters.

Breaking of these chiral symmetries is sufficiently small 
because the ratios of SM fermion masses $m_{f}$ for a given renormalization scale are 
about $O(10^{-2})$ \cite{Xing:2011aa}, 
\begin{align}
{m_{u} \over  m_{c}} \sim {1\over 500} \, ,  ~~
{m_{d} \over m_{s}} \sim {1\over 20} \, , ~~
{m_{e} \over m_{\mu}} \sim {1\over 200} \, . 
\end{align}
With some grand unified theories in mind, 
it is natural to assume that the Dirac mass matrix $m_{D}$ also has such approximate chiral symmetries. 
Thus, in this letter, we perform a chiral perturbative analysis \cite{Gasser:1984gg}
by the smallest singular value $m_{D1}$ of $m_{D}$ in
the type-I seesaw mechanism \cite{Minkowski:1977sc,GellMann:1980v,Yanagida:1979as, Mohapatra:1979ia}. \\

\section{chiral perturbative analysis by $m_{D1}$}

For the singular value decomposition (SVD) $m_{D} = U_{D} m_{D}^{\rm diag} V_{D}^{T}$, 
it does not lose generality to consider the diagonal basis of $m_{D}$ 
by field redefinitions of unitary matrices $U_{D}$ and $V_{D}$ \cite{Endoh:2002wm}. 
If the mass matrix $m_{\n}$ of left-handed neutrinos $\n_{Li}$  is regular and invertible, 
the mass matrix $M_{R}$ of right-handed neutrinos $\n_{Ri}$  in the type-I seesaw mechanism is reconstructed as
\begin{align}
M_{R} = m_{D}^{T} m_{\n}^{-1} m_{D} 
= 
\Diag{m_{D1}}{m_{D2}}{m_{D3}} 
m_{\n}^{-1}
\Diag{m_{D1}}{m_{D2}}{m_{D3}} ,
\label{4}
\end{align}
where $m_{Di}$ are the singular values of $m_{D}$. 

In the limit of $m_{D1} \to 0$, $m_{D}$ and $M_{R}$ have $U(1)_{1R}$ chiral symmetry associated with the lepton number of $\n_{R1}$ \cite{Branco:1988ex, Adhikari:2010yt}; 
\begin{align}
m_{D} \, R (\th_{R}) = m_{D} \, , ~~
M_{R} \, R (\th_{R}) = M_{R} \, . 
\end{align}
These conditions are equivalent to $m_{D}$ and $M_{R}$ having no eigenvector components in the massless modes \cite{Yang:2022bex}; 
\begin{align}
m_{D} \Column{1}{0}{0} = M_{R} \Column{1}{0}{0} = \Column{0}{0}{0} \, . 
\end{align}

Conversely, if $M_{R}$ has a certain chiral symmetry, 
$m_{D}$ has the same symmetry \cite{Yang:2022bex}.
The proof is as follows. 
SVDs of two matrices (with rank two) $M_{R} = V M_{R}^{\rm diag} V^{T}$ and $m_{D} = U_{D} m_{D}^{\rm diag} V_{D}^{T}$ lead to 
\begin{align}
V \Diag{0}{M_{2}}{M_{3}} V^{T} 
=  V_{D} \Diag{0}{m_{D2}}{m_{D3}} U_{D}^{T}
m_{\n}^{-1}
 U_{D} \Diag{0}{m_{D2}}{m_{D3}} V_{D}^{T} 
 \, , \label{7}
\end{align}
where $M_{i}$ are singular values of $M_{R}$. 
By performing production of matrices between two $m_{D}^{\rm diag}$, 
\begin{align}
\Diag{0}{M_{2}}{M_{3}}
= V^{\dg} V_{D} 
\begin{pmatrix}
0 & 0& 0 \\
0 & * & * \\
0 & * & *
\end{pmatrix}
 V_{D}^{T} V^{*} \, ,
\end{align}
where $*$ denotes any matrix element. 
Since this is also a SVD of $M_{R}$, 
$V^{\dg} V_{D}$ must be a unitary matrix in the 2-3 subspace;
\begin{align}
V^{\dg} V_{D} = 
\begin{pmatrix}
1 & 0 & 0 \\
0 & * & * \\
0 & * & * \\
\end{pmatrix} . 
\end{align}
Therefore, the first eigenvector of  $V$ and $V_{D}$ coincide in this limit, and the two mass matrices share the same chiral symmetry.
Of course this $U(1)_{1R}$ symmetry must be broken because the massless $\n_{R1}$  contradicts observations.  

On the other hand, it appears unreasonable that the kernels of $m_{D}$ and $M_{R}$, which can be given arbitrarily in a  model, must coincide. This point can rather be considered as a constraint on the seesaw mechanism.
In the basis where $M_{R}$ is diagonal, 
a parameterization of $m_{D}$ 
\begin{align}
m_{D} =
\begin{pmatrix}
A_1 & B_1 & C_1 \\
A_2 & B_2 & C_2 \\
A_3 & B_3 & C_3 \\
\end{pmatrix} 
\equiv (\bm A \, , \bm B \, , \bm C) \, , 
\end{align}
yields the natural representation~\cite{Barger:2003gt} of $m_{\n}$ given by
\begin{align}
m_{\n} &= m_{D} M_{R}^{-1} m_{D}^{T} 
=  {1 \over M_{1}} {\bm A} \otimes {\bm A}^{T} + {1 \over  M_{2}}  \bm B \otimes \bm B^{T} + {1 \over M_{3}} \bm C \otimes \bm C^{T} \, . 
\label{formula}
\end{align}
If the kernels of $m_{D}$ and $M_{R}$ do not coincide, such as limits $A_{i} \to 0$ and $M_{2} \to 0$, the matrix $m_{\n}$ have two extremely different mass scales. 
If we expect $m_{\n}$ to be non-hierarchical,
magnitudes of $A_{i} A_{j} /M_{1}$ and $B_{i} B_{j}/M_{2}$ must be comparable. 
This means that the chiral symmetries associated with the first generation must coincide, 
and moreover, its breaking parameter $\e_{R}$ is common to some extent; 
\begin{align}
{A_{i}^{2} \over B_{i}^{2}}  \sim  {M_{1} \over M_{2} } \sim \e_{R}^{2} \, . 
\label{Rchiral}
\end{align}
%


Next we analyze the mass matrix $M_{R}$~(\ref{4})
treating $m_{D1}$ as a perturbation. 
For $(M_{R})_{ij} = m_{Di} (m_{\n}^{-1})_{ij} m_{Dj}$, 
let $M_{R0}$ be the unperturbed mass matrix with $m_{D1} = 0$.
Clearly matrix elements of $M_{R0}$ are limited to the 2-3 submatrix, 
and a unitary matrix $V_{0}$ that diagonalizes $M_{R0}$ rotates the subspace.

The full matrix $M_{R}$ in the diagonalized basis of $M_{R0}$ is
\begin{align}
V_{0}^{\dg} M_{R} V_{0}^{*} &= 
\begin{pmatrix}
(\d^{2} M_R)_{11} & (\d M_R)_{12} & (\d M_R)_{13} \\
(\d M_R)_{12} & M_{2}^{(0)} & 0  \\
(\d M_R)_{13} & 0 & M_{3}^{(0)}
\end{pmatrix}  \equiv  M' \, ,  \label{VMR}
\end{align}
where $M_{i}^{(0)}$ are the singular values of $M_{R0}$ without perturbation and 
\begin{align}
(\d^{2} M_R)_{11} & = m_{D1}^{2} (m_{\n}^{-1})_{11} \, ,  \\
(\d M_R)_{1(2,3)} & =  m_{D1}  (m_{\n}^{-1})_{12} m_{D2} (V_{0})_{2(2,3)}^{*} + m_{D1} (m_{\n}^{-1})_{13} m_{D3} (V_{0})_{3(2,3)}^{*} \, . 
\end{align}
From these expressions, corrections of the diagonalization occur in the first order of $m_{D1}$, and the lightest mass $M_{1}$ does in the second order. 
As long as the smallness of perturbation $m_{D1}$ allows the approximation $M_{2,3} \simeq M_{2,3}^{(0)}$, 
the diagonalization of Eq.~(\ref{VMR}) is evaluated in a similar way to the seesaw mechanism,
\begin{align}
M_{1} \simeq \abs {(\d^{2} M_R)_{11}- {(\d M_R)_{12}^{2} \over M_{2}^{(0)}} - {(\d M_R)_{13}^{2} \over M_{3}^{(0)}} }\, .  
\label{M1}
\end{align}
This is consistent with a formal solution of perturbed SVD to the second order
\cite{Yang:2022bex}. 
Although this expression diverges in the limit of $M_{2,3}^{(0)} \to 0$, 
the approximation $M_{2,3}^{(0)} \simeq M_{2,3}$ does not hold in this situation, 
and the diagonalization must be done correctly. 
The applicability of this perturbation theory will be discussed later. 

This result can also be considered from another viewpoint. 
Multiplying Eq.~(\ref{M1}) by $M_{2}^{(0)} M_{3}^{(0)}$ leads to 
\begin{align}
 M_{1} M_{2}^{(0)} M_{3}^{(0)} \simeq  \abs{ M'_{11} M'_{22} M'_{33} - M'_{12} M'_{21} M'_{33} - M'_{13} M'_{31} M'_{22} } = | \Det M' |  = |\Det M_{R}| \, . 
\end{align}
The determinant and 
the minor determinant $\det M_{R0}$ which is restricted to the heavier two generations 
are,
\begin{align}
|\Det M_{R} | & = M_{1} M_{2} M_{3} = \Pi_{i = 1}^{3} m_{Di}^{2} \, | \Det m_{\n}^{-1}| \, , \\
|\det M_{R 0 } | & = M_{2}^{(0)} M_{3}^{(0)} = \Pi_{i=1}^{2} m_{Di}^{2} \, | \det  m_{\n}^{-1}| \, . 
\end{align}
%
The cofactor of the inverse matrix $\det m_{\n}^{-1} = (m_\n)_{11} / \Det m_{\n}$ yields
\begin{align}
M_{1} \simeq \abs{\Det M_{R} \over \det M_{R 0}} = m_{D1}^{2} \abs{\Det m_{\n}^{-1} \over \det m_{\n}^{-1} }
= { m_{D1}^{2} \over | (m_\n)_{11} | } \, .
\label{relation0}
\end{align}
Therefore, the lightest mass $M_{1}$ is expressed by the matrix element of $m_{\n}$ in the basis where $m_{D}$ is diagonal. 
The right-hand side contains a basis-dependent quantity, 
because the minor determinant is evaluated in a particular basis.

This relation is equivalent to the result of integrating out heavy neutrinos by $\del \Lg / \del \n_{R 2,3} = 0$. 
Removing the heavy generations like the seesaw mechanism, we obtain
\begin{align}
M_{1} &= \abs {(M_{R})_{11} - \sum_{\a, \b =2}^{3} (M_{R})_{1 \a} (M_{R0}^{-1})_{\a \b} (M_{R})_{\b 1}  } \\
 &= m_{D1}^{2} \abs{ (m_{\n}^{-1})_{11} - \sum_{\a, \b =2}^{3} (m_{\n}^{-1})_{1 \a} (m_{\n}^{-1})_{\a\b}^{-1} (m_{\n}^{-1})_{\b 1} } = { m_{D1}^{2} \over \abs{(m_{\n})_{11}} }\, ,
\end{align}
where $(m_{\n}^{-1})_{\a \b}^{-1} $ denotes an inverse matrix of $(m_{\n}^{-1})$ when it is restricted to the 2-3 submatrix. 
By the simple normalization $m_{D1} \sim m_{u,d,e} \sim 1 \, \MeV$ and $|(m_{\n})_{11}| \sim 1 \, \meV$, 
\begin{align}
M_{1} \simeq \lsp {m_{D1} \over 1 \, \MeV }\rsp^{2} \lsp {1 \, \meV \over | (m_\n)_{11} | } \rsp \, 10^{6} \, \GeV \, . 
\label{relation1}
\end{align}
%

\section{Applicable limit of perturbativity} 

This mass relation diverges in the limit of $(m_{\n})_{11} \to 0$, indicating that the perturbation theory is not valid for too small $(m_{\n})_{11}$.
Since this limit corresponds to $\det m_{\n}^{-1} = 0$ and $M_{2}^{(0)}$(or $M_{3}^{(0)}$) $= 0$, the second (or third) term in Eq.~(\ref{M1}) causes the divergence. 
First, let us consider the applicability of the chiral perturbation theory from general observation. 
By using dimensionless parameters $\e_{L,R}$ representing the chiral symmetry breaking of $m_{\n}$ and $M_{R}$, Eq.~(\ref{4}) is  denoted as  
\begin{align}
M_{R} = 
\begin{pmatrix}
\e_{R}^{2} & \e_{R} & \e_{R} \\
\e_{R} & * & * \\
\e_{R} & * & * 
\end{pmatrix}
= 
\Diag{m_{D1}}{m_{D2}}{m_{D3}} 
\begin{pmatrix}
\e_{L}^{2} & \e_{L} & \e_{L} \\
\e_{L} & * & *   \\
\e_{L} & * & *   \\
\end{pmatrix}^{-1}
\Diag{m_{D1}}{m_{D2}}{m_{D3}} ,
\end{align}
where $O(1)$ coefficients are omitted. 
Because of the constraint $\e_{R} \, \e_{L} \propto m_{D1}$, 
the smaller breaking parameter $\e_{L}$ yields the larger $\e_{R}$. 
Therefore, for the approximate chiral symmetry of $M_{R}$ to be a good description, 
the smallness of $m_{D1}$ must not be cancelled by $m_{\n}^{-1}$.  
This property is due to the commutativity of the left and right chiral transformations
 in the diagonalized basis of $m_{D}$.

More specifically, this validity of the perturbation theory can be shown as conditions on the matrix elements. To this end,  
we examine another mass relation of $M_{2}$
by a similar perturbative analysis for $m_{D2} \ll m_{D3}$; 
\begin{align}
M_{2} \simeq  {|\det m_{D}|^{2}  |\det m_{\n}^{-1}| \over m_{D3}^{2} |(m_{\n}^{-1})_{33}|} 
=  m_{D2}^{2}  \abs{\det m_{\n}^{-1} \over (m_{\n}^{-1})_{33}} 
=  m_{D2}^{2}  \abs{(m_{\n})_{11} \over \det' m_{\n}} 
\, , 
\end{align}
where $\det'$  denotes a minor determinant restricted to 1-2 submatrix. 
A normalization for $m_{D2} \sim 100 \, \MeV$ leads to 
\begin{align}
M_{2} \sim \lsp {m_{D2} \over 100 \, \MeV} \rsp^{2} 
 \abs{(m_{\n})_{11} \, 10 \, \meV \over \det' m_{\n}}  10^{9} \, \GeV \,  . 
\label{22}
\end{align}
Indeed, the limit of $m_{11} \to 0$ yields $M_{2} \to 0$ and breaks the perturbation theory (of the first generation). 
Conditions to prevent such a breakdown are
\begin{align}
 {M_{1} \over M_{2} } \simeq
{m_{D1}^{2} \over  m_{D2}^{2} } \abs{  \det' m_{\n} \over (m_\n)_{11}^{2}  } \lesssim 0.1  \, , ~~~ 
{ m_{D1}  \over m_{D2}  } \lesssim 0.3 \abs{(m_\n)_{11} \over \sqrt{ \det' m_{\n}} } \, .
\label{23}
\end{align}

In order to make $M_{1}$ larger, 
$(m_{\n})_{11}$ and $\det' m_{\n}$  in Eq.~(\ref{23}) must be small simultaneously. 
In this case, the 1-2 submatrix of $m_{\n}$ approaches singular;
\begin{align}
m_{\n} \sim (m_{\n})_{22}
\begin{pmatrix}
\e^{2} & \e & *  \\
\e & 1 & * \\ 
* & * & *
\end{pmatrix} ,  
\end{align}
where $\e \sim  \abs{(m_\n)_{11} / \sqrt{ \det' m_{\n}} }$ is a small dimensionless parameter and $O(1)$ coefficients are ignored.  
Since the perturbability for $M_{1}/M_{3}$ provides a similar restriction for the 1-3 element, 
a heavy $M_{1}$ requires that $m_{\n}^{-1}$ compensates for the smallness of $m_{D1}$ as
\begin{align}
m_{\n}^{-1}  \sim 
\begin{pmatrix}
\e^{-2} & \e^{-1} & \d^{-1} \\
\e^{-1} & * & * \\
\d^{-1} & * & *
\end{pmatrix} \, ,  ~~~ \e \gtrsim {m_{D1} \over m_{D2} } \, , ~~ \d \gtrsim {m_{D1} \over m_{D3} } \, , 
\end{align}
where $\d$ is another small parameter.
Eventually, $M_{1}$ will only be heavy when approximate chiral symmetries of $m_{D}$ and $m_{\n}$ are almost identical;
\begin{align}
R(\th_{L} ) \,  m_{\n} \simeq m_{\n} \, , ~~~
R(\th_{L}) \, m_{D} \simeq m_{D} \, .
\label{Lchiral}
\end{align}
If the breaking parameter $\e \, (\d)$ is smaller than $m_{D1}/m_{D2 \, (3)}$, 
the perturbation theory for $M_{1}/ M_{2  \, (3)}$ is no longer valid.
This feature is similar to the discussion around Eq.~(\ref{Rchiral}), and 
it indicates that $M_{R}$ and/or $m_{\n}$ share the chiral symmetry associated with $m_{D1}$.

Moreover, Eq.~(\ref{Lchiral}) requires the zero eigenvector $\bm v_{0}$ of $m_{D}$
is close to that of $m_{\n}$. 
In the basis where  $m_{\n}$ is diagonalized by the MNS matrix, $\bm v_{0}$ is close to $(0 \, ,1 \, , -1)$ or $(-2 \, ,1 \, ,1)$ for the normal hierarchy (NH) or the inverted hierarchy (IH).
Since this is equivalent to $m_{D}$ having no $\bm v_{0}$ component,  
the form of $m_{D}$ in the case of NH is 
\begin{align}
m_{D} \simeq 
\begin{pmatrix}
A_{2} & B_{2} & C_{2} \\
A_{2} & B_{2} & C_{2} \\
A_{2} & B_{2} & C_{2} \\
\end{pmatrix}
+ 
\begin{pmatrix}
0 & 0 & 0 \\
A_{3} &B_{3} &C_{3} \\
-A_{3} & - B_{3} & - C_{3} \\
\end{pmatrix} . 
\end{align}
Thus, it is difficult to impose a strong hierarchy of $|(m_{D})_{33}| \gg |(m_{D})_{ij}|$.

\vspace{12pt}

The mass relation~(\ref{relation1}) leads to significant phenomenological consequences. 
\begin{enumerate}
\item 
For the thermal leptogenesis \cite{Fukugita:1986hr} by $\n_{R1}$,  
the famous lower limit of the mass $M_{1} \gtrsim 10^{9} \, \GeV$ \cite{Davidson:2002qv, Hamaguchi:2001gw} implies that
\begin{align}
m_{D1} \gtrsim 30 \, \MeV \, . 
\end{align}
Therefore, except in a special condition that amplifies the mass $M_{1}$,  
a simple unified theory with $m_{D1} \sim 1 \, \MeV$ and the type-I seesaw mechanism 
seems to be difficult to reconcile with the thermal leptogenesis by $\n_{R1}$. 
This feature is expected in wide parameter regions of many models\footnote{Note that such a bound is inconsistent with the existence of long-lived particles \cite{Khlopov:1984pf}, and it does not immediately rule out the possibility of leptogenesis. }.

\item 
If diagonalization of the Yukawa matrices of leptons $Y_{\nu, e}$ has only small mixings, 
the value $(m_{\n})_{11}$ is close to the effective mass $m_{ee}$ of the neutrinoless double beta decay \cite{Vergados:1985pq}.  
Although NH has a canceling region $m_{ee} \simeq 0$, there is no chiral symmetry because $m_{1} \sim 3$ meV, and the chiral perturbation theory simply breaks down. 
Since the lepton mass is not susceptible to renormalization, 
$m_{D1}$ is expected to be about $m_{D1} \simeq 0.5 \MeV$ 
from singular values at the GUT scale $m_{u} \simeq m_{e} \simeq 0.5 \MeV$. 
Therefore, the lightest mass is about $M_{1} \gtrsim O(100) \, \TeV$ in NH and $M_{1} \sim O(10) \, \TeV$ in IH. 

\item 
Although the lightest TeV-scale right-handed neutrino $\n_{R1}$ only has a tiny Yukawa coupling $y_{\n1} \sim O(10^{-5})$ and a very weak Higgs interaction, 
such a $\n_{R1}$ may be indirectly involved in the anomaly called IceCube gap \cite{IceCube:2014stg}. 

\end{enumerate}

If $m_{\n}$ has a good chiral symmetry, this relation can be applied for $m_{\n}$. 
A similar discussion for the seesaw formula $m_{\n} = m_{D}^{\rm diag} M_{R}^{-1} m_{D}^{\rm diag}$ and perturbatively small $m_{D1}$ leads to ,
\begin{align}
m_{1} =  \abs{\Det m_{\n} \over \det m_{\n 0 }} = m_{D1}^{2} \abs{\Det M_{R}^{-1} \over \det M_{R}^{-1} }
= { m_{D1}^{2} \over | (M_{R})_{11} | }  \, ,  
\label{25}
\end{align}
and $| (M_{R})_{11} | \sim 1\, \PeV$ holds for $m_{1} \sim 1 \, \meV$. 
However, we need to be careful about this argument. 
For a strongly hierarchical $M_{R}$ with $M_{3} \gg M_{1,2}$, 
the lightest mass $m_{1}$ comes from $1/M_{3}$ by the sequential dominance \cite{King:1999cm} and Eq.~(\ref{25}) is not the correct relationship.
Since $M_{R}$ seems to be much closer to a singular matrix than $m_{\n}$,
the mass relation appears safer to consider only for $m_{\n}^{-1}$.  

In both cases of NH and IH, 
the eigenvectors associated with the lightest mass $m_{1 \, \rm or \, 3}$ are not in the direction $(1,0,0)$. Since chiral symmetries of left-handed fields seem not to be shared between $m_{D}$ and $m_{\n}$, it is natural to think that the smallness of $m_{D1}$ rather ensures the hierarchy of $m_{2} /m_{3}$ in NH.

Finally, this mass relation must be almost stable against quantum corrections 
 because it is associated with the approximate chiral symmetry  \cite{tHooft:1979bh} of the right-handed neutrino $\n_{R1}$ and the gauge charges of SM cancel out between $m_{\n}$ and $m_{D}$. Thus, this is considered a general constraint on the type-I seesaw mechanism.

\section{Summary}

In this letter, we perform a perturbative analysis by the lightest singular value $m_{D1}$ of the Dirac mass matrix $m_{D}$ in the type-I seesaw mechanism.
The lightest mass $M_{1}$ of the right-handed neutrino $\nu_{R1}$ is expressed as $M_{1} = m_{D1}^{2}/ |(m_{\nu})_{11}|$ by the mass matrix of the left-handed neutrinos $m_{\nu}$ in the diagonal basis of $m_{D}$. 
A similar relationship $M_{2} \propto m_{D2}$ is also obtained for the second generation.

This chiral perturbation theory breaks down when $m_{\n}^{-1}$ cancels the hierarchy of $m_{D}$ and it corresponds to a situation where an approximate chiral symmetry $m_{\nu}$ and $m_{D}$  for the left-hand field $\n_{L1}$ is almost identical. 

Since $m_{D1} \sim 0.5 \, \MeV$ leads to $M_{1} \gtrsim O(100) \, \TeV$ for NH and $M_{1} \sim O(10) \, \TeV$ for IH, such a light TeV-scale right-handed neutrino 
with a tiny Yukawa coupling of $y_{\n1} \sim O(10^{-5})$ can indirectly influence various observations.
On the other hand, the famous bound of the thermal leptogenesis $M_{1} \gtrsim 10^{9}$ GeV that requires $m_{D1} \gtrsim 30$ MeV seems to be difficult to reconcile with a simple  unified theory without a special condition.


\begin{thebibliography}{10}

\bibitem{Weinberg:1968de}
S.~Weinberg,
\newblock Phys. Rev. {\bf 166}, 1568 (1968).

\bibitem{Witten:1978qu}
E.~Witten,
\newblock Nucl. Phys. B {\bf 145}, 110 (1978).

\bibitem{tHooft:1979bh}
G.~'t~Hooft,
\newblock NATO Adv.Study Inst.Ser.B Phys. {\bf 59}, 135 (1980).

\bibitem{Xing:2011aa}
Z.-z. Xing, H.~Zhang, and S.~Zhou,
\newblock Phys. Rev. D {\bf 86}, 013013 (2012), arXiv:1112.3112.

\bibitem{Gasser:1984gg}
J.~Gasser and H.~Leutwyler,
\newblock Nucl. Phys. B {\bf 250}, 465 (1985).

\bibitem{Minkowski:1977sc}
P.~Minkowski,
\newblock Phys. Lett. {\bf 67B}, 421 (1977).

\bibitem{GellMann:1980v}
M.~Gell-Mann, P.~Ramond, and R.~Slansky,
\newblock Conf. Proc. {\bf C790927}, 315 (1979).

\bibitem{Yanagida:1979as}
T.~Yanagida,
\newblock Conf. Proc. {\bf C7902131}, 95 (1979).

\bibitem{Mohapatra:1979ia}
R.~N. Mohapatra and G.~Senjanovic,
\newblock Phys. Rev. Lett. {\bf 44}, 912 (1980).

\bibitem{Endoh:2002wm}
T.~Endoh, S.~Kaneko, S.~K. Kang, T.~Morozumi, and M.~Tanimoto,
\newblock Phys. Rev. Lett. {\bf 89}, 231601 (2002), arXiv:hep-ph/0209020.

\bibitem{Branco:1988ex}
G.~C. Branco, W.~Grimus, and L.~Lavoura,
\newblock Nucl. Phys. B {\bf 312}, 492 (1989).

\bibitem{Adhikari:2010yt}
R.~Adhikari and A.~Raychaudhuri,
\newblock Phys. Rev. D {\bf 84}, 033002 (2011), arXiv:1004.5111.

\bibitem{Yang:2022bex}
M.~J.~S. Yang,
\newblock (2022), arXiv:2211.15101.

\bibitem{Barger:2003gt}
V.~Barger, D.~A. Dicus, H.-J. He, and T.-j. Li,
\newblock Phys. Lett. B {\bf 583}, 173 (2004), arXiv:hep-ph/0310278.

\bibitem{Fukugita:1986hr}
M.~Fukugita and T.~Yanagida,
\newblock Phys. Lett. {\bf B174}, 45 (1986).

\bibitem{Davidson:2002qv}
S.~Davidson and A.~Ibarra,
\newblock Phys. Lett. {\bf B535}, 25 (2002), arXiv:hep-ph/0202239.

\bibitem{Hamaguchi:2001gw}
K.~Hamaguchi, H.~Murayama, and T.~Yanagida,
\newblock Phys. Rev. {\bf D65}, 043512 (2002), arXiv:hep-ph/0109030.

\bibitem{Khlopov:1984pf}
M.~Y. Khlopov and A.~D. Linde,
\newblock Phys. Lett. B {\bf 138}, 265 (1984).

\bibitem{Vergados:1985pq}
J.~D. Vergados,
\newblock Phys. Rept. {\bf 133}, 1 (1986).

\bibitem{IceCube:2014stg}
IceCube, M.~G. Aartsen {\em et~al.},
\newblock Phys. Rev. Lett. {\bf 113}, 101101 (2014), arXiv:1405.5303.

\bibitem{King:1999cm}
S.~F. King,
\newblock Nucl. Phys. B {\bf 562}, 57 (1999), arXiv:hep-ph/9904210.

\end{thebibliography}

\end{document}